\documentclass[10pt]{article}

 \usepackage{amsmath,amsthm,amssymb}
 \usepackage{makeidx,epsfig,lscape}
 \usepackage{color,colortbl}
 \usepackage{fancyhdr}
 \usepackage{xcolor,pict2e}
 \usepackage{multirow}
 \renewcommand{\headrulewidth}{0pt}
 \renewcommand{\footrulewidth}{0.5pt}

 \definecolor{myaqua}{rgb}{0.0,0.5,0.55}
 \definecolor{lightaqua}{rgb}{0.75,0.95,0.95}

 \usepackage[colorlinks = true,
            linkcolor = myaqua,
            urlcolor  = blue,
            citecolor = myaqua]{hyperref}

\usepackage{caption}
\usepackage{floatrow}
\def\lin#1#2{\textcolor[rgb]{0.6,0.6,0.6}{\vspace*{#1mm} \hrule
   height 3 pt \vspace*{#2mm}}}
\def\bt{\begin{tabular}}
\def\et{\end{tabular}}
\def\and{\mbox{ and }}

\def\1{{\bf 1}}

 \def\sectionn#1{\refstepcounter{section}{\color{myaqua}

 \vskip 6mm

 \noindent\Large\bf\thesection. #1}

 \vskip 3mm}

 \def\boxx#1#2#3#4#5{
 {\linethickness{#4pt}\put(#1,#5){\color{myaqua}{\line(1,0){#3}}}}
 \multiput(#1,#2)(0,#4){2}{\line(1,0){#3}}
 \multiput(#1,#2)(#3,0){2}{\line(0,1){#4}}
  }

\begin{document}

 \fancyhead[L]{\hspace*{-13mm}
 \bt{l}{\bf Open Journal of *****, 2020, *,**}\\
 Published Online **** 2020 in SciRes.
 \href{http://www.scirp.org/journal/*****}{\color{blue}{\underline{\smash{http://www.scirp.org/journal/****}}}} \\
 \href{http://dx.doi.org/10.4236/****.2020.*****}{\color{blue}{\underline{\smash{http://dx.doi.org/10.4236/****.2014.*****}}}} \\
 \et}
 \fancyhead[R]{\includegraphics{pic1.ps}}

 $\mbox{ }$

 \vskip 12mm

{ 

{\noindent{\huge\bf\color{myaqua}
  Momentum eigensolutions of Feinberg-Horodecki \\[2mm] equation with time-dependent screened Kratzer-\\[2mm] Hellmann potential}}
%
\\[6mm]
{\large\bf Mahmoud Farout$^{1}$, Sameer M. Ikhdair$^{1,2}$}}
\\[2mm]
{ 
 $^1$ Department of Physics, An-Najah National University, Nablus, Palestine\\
Email: \href{mailto:m.qaroot@najah.edu}{\color{blue}{\underline{\smash{m.qaroot@najah.edu}}}}\\[1mm]
$^2$ Department of Electrical Engineering, Near East University, Nicosia, Northern Cyprus, Mersin 10, Turkey\\
Email:
\href{mailto:sameer.ikhdair@najah.edu}{\color{blue}{\underline{\smash{sameer.ikhdair@najah.edu}}}}
 \\[4mm]
Received June, 6, 2020
 \\[4mm]
Copyright \copyright \ 2020 by author(s) and Scientific Research Publishing Inc. \\
This work is licensed under the Creative Commons Attribution International License (CC BY). \\
\href{http://creativecommons.org/licenses/by/4.0/}{\color{blue}{\underline{\smash{http://creativecommons.org/licenses/by/4.0/}}}}\\
 \includegraphics{pic2.ps}

\lin{5}{7}

 { 
 {\noindent{\large\bf\color{myaqua} Abstract}{\bf \\[3mm]
 \textup{We obtain an approximate value of the quantized momentum eigenvalues, $P_n$, together with the space-like coherent eigenvectors for the space-like counterpart of the Schrödinger equation, the Feinberg-Horodecki equation, with a screened Kratzer-Hellmann potential which is constructed by the temporal counterpart of the spatial form of this potential. In addition, we got exact eigenvalues of the momentum and the eigenstates by solving Feinberg-Horodecki equation with Kratzer potential. The present work is illustrated with  three special cases of the screened Kratzer-Hellman potential: the time-dependent screened Kratzer potential, time-dependent Hellmann potential and, the time-dependent screened Coulomb potential.
 }}}
 \\[4mm]
 {\noindent{\large\bf\color{myaqua} Keywords}{\bf \\[3mm]
 Coherent states; Feinberg-Horodecki equation; Time-dependent screened Kratzer-Hellmann potential; Time-dependent Kratzer potential; Time-dependent Yukawa potential; Time-dependent Hellmann potential; and time-dependent Coulomb potential
}

 \fancyfoot[L]{{\noindent{\color{myaqua}{\bf How to cite this
 paper:}} M. Farout and S. M. Ikhdair (2020)
Momentum eigensolutions of Feinberg-Horodecki equation with time-dependent screened Kratzer-Hellmann potential.
 ***********,*,***-***}}

\lin{3}{1}

\sectionn{Introduction}

{ \fontfamily{times}\selectfont
 \noindent Momentum is one notion of the most unmeasured physical quantity. The scientific meaning and the conceptual generality of momentum only appeared after extensive efforts were made to calculate it since the beginning of the $19^{ th}$ century. That work culminated lately in deriving a second-order differential equation which has shown the recent essential status momentum has in quantum mechanics. Nowadays, momentum calculations are pushing these foundations to new extremes by probing smaller energies and smaller length scales, where quantum mechanics appears to play a dominating role.
 
 Any physical phenomenon in nature is usually characterized by solving second-order differential equations. The time-dependent Schrödinger equation is considered as an example that describes quantum-mechanical phenomena, in which it dictates the dynamics of a quantum system. The solution of this differential equation is carried out by means of any method results in the eigenvalues and eigenfunctions of that quantum system. However, the solution of time-dependent Schrödinger equation analytically is not easy except when the time-dependent potentials are constant, linear, and quadratic functions of the coordinates \cite{Park02, Vorobeichik98, Shen03, Feng01}.
 The Feinberg-Horodecki (FH) equation is a space-like counterpart of the Schrödinger equation which was found by Horodecki \cite{Horodecki88} from the relativistic Feinberg equation \cite{Feinberg67}. 
 The space-like solutions of the (FH) equation can be employed to test its relevance in different areas of sciences including physics, biology and medicine \cite{Molski06, Molski10, Witten81}. Molski constructed the space-like bound states of a time-dependent form of Morse oscillator \cite{Molski06} and anharmonic oscillators \cite{Molski10} on the basis of the FH equation to minimize the uncertainty in the time-energy relation and showed that the results are useful for interpreting the formation of the specific growth patterns during crystallization process and biological growth. Hamzavi et al \cite{Hamzavi13} obtained the exact bound state solutions of the FH equation with rotating time-dependent Deng-Fan oscillator potential by using parametric Nikiforov–Uvarov (NU) method. Moreover, Eshghi et al \cite{Eshghi16} solved FH equation for time-dependent mass distribution (TDM) harmonic oscillator quantum system with a certain interaction applied to a mass distribution m(t) to provide a particular spectrum of stationary energies. Also, the spectrum of harmonic oscillator potential $V(t)$ acting on TDM $m(t)$ oscillator was obtained. 
 
 In a non-relativistic case, the NU method was employed to obtain the bound state solutions of arbitrary angular momentum Schrödinger equation with the modified Kratzer potential which was also widely used in studying atomic physics, molecular physics and quantum chemistry \cite{Berkdemir06}. On the other hand, the factorization method was used to obtain the solution of the non-central modified Kratzer potential for the diatomic molecules \cite{Sadeghi07}. The exact solutions of the Schrödinger equation with modified Kratzer and corrected Morse potentials with position-dependent mass were also obtained \cite{Sever08}. A particle's coherent states in Kratzer potentials are constructed by solving Feynman’s path integral \cite{Kandirmaz18}.  Further, the exact solution of the Schrödinger equation for the modified Kratzer potential plus a ring-shaped potential was solved \cite{Cheng07exact}.
 
 Besides, in the relativistic scale, approximate solutions of the D-dimensional Klein-Gordon equation are obtained for the scalar and vector general Kratzer potential for any $l$ by using the ansatz method and the solutions of the Dirac equation with equal scalar and vector ring-shaped modified Kratzer potential were found by means of the NU method \cite{Cheng07Solution,Hassanabadi11}. 
 
 At the level of applications, many authors have studied the modified Morse-Kratzer potential for alkali hydrides \cite{Ghodgaonkar81}, and the effect of modified Kratzer potential on the confinement of an exciton in a quantum dot \cite{Khordad14}. Further, an analysis of the applications of the modified Kratzer potential, the bound states of two special cases of the interactions and this approximation was used to obtain the solution of the Schrödinger equation for the Morse potential \cite{Babaei-Brojeny11}.
 
 Recently, a superposition of modified Kratzer potential plus screened Coulomb potential was suggested to study diatomic molecules \cite{Edet20}.  Edet et al have obtained an approximate solution of the Schrödinger equation for the modified Kratzer potential plus screened Coulomb potential model, within the framework of NU method. They obtained bound state energy eigenvalues for $N_2$, $CO$, $NO$, and $CH$ diatomic molecules for various vibrational and rotational quantum numbers. Three special cases were considered by changing the potential parameters, resulting in the form of the modified Kratzer potential, the screened Coulomb potential, and the standard Coulomb potential \cite{Edet20}. Further, Okorie et al have solved the Schrödinger equation with the modified Kratzer plus screened Coulomb potential using the modified factorization method. They have used an approximation proposed by Greene–Aldrich with a suitable transformation scheme to obtain the energy eigenvalues equation and the corresponding energy eigenstates for $CO$, $NO$, and $N_2$ diatomic molecules. They have used the energy eigenvalues of the modified Kratzer plus screened Coulomb potential to obtain the vibrational partition functions and other thermodynamic functions for the selected diatomic molecules \cite{Okorie20}. 
 
 In a new work, we have obtained the quantized momentum solution of the FH equation with combined Kratzer plus screened Coulomb potential using NU method. We constructed three special cases of this general form; the time-dependent modified Kratzer potential, the time-dependent screened Coulomb potential and the time-dependent Coulomb potential \cite{Farout20}.
 
 Very recently, the Hellmann potential \cite{Hellmann35, Hellmann36} is considered as a combination of Coulomb plus Yukawa potentials \cite{Osobonye20}. In this regard, many authors have extended their works to study the screened Kratzer and Hellmann potential (SKHP). Osobonye et al \cite{Osobonye20} have obtained the eigenvalues and eigenfunctions of the Schrödinger equation with newly proposed SKHP via the conventional NU method. 
 
 The aim of this work is to apply the NU method \cite{Nikiforov88} for a system of screened Kratzer potential plus Hellmann potential having a certain time-dependence. The momentum eigenvalues, $P_n$, of the FH equation and the space-like coherent eigenvectors are obtained. The rest of the present work is organized as follows: the NU method is briefly reviewed in Section 2. An approximate eigensolution of the FH equation with the time-dependent screened Kratzer and Hellmann potential (SKHP) model is given in Section 3. We calculate the momentum states together with eigenvectors of FH equation.
 Further, the solutions of a few special potentials which are mainly found from our general molecular solution are presented in section 4. Finally, we give our conclusions in section 5.

\renewcommand{\headrulewidth}{0.5pt}
\renewcommand{\footrulewidth}{0pt}

 \pagestyle{fancy}
 \fancyfoot{}
 \fancyhead{} 
 \fancyhf{}
 \fancyhead[RO]{\leavevmode \put(-90,0){\color{myaqua}M. Farout, S. M. Ikhdair} \boxx{22}{-10}{10}{50}{15} }
 \fancyhead[LE]{\leavevmode \put(0,0){\color{myaqua}M. Farout, S. M. Ikhdair}  \boxx{-45}{-10}{10}{50}{15} }
 \fancyfoot[C]{\leavevmode
 \put(0,0){\color{lightaqua}\circle*{34}}
 \put(0,0){\color{myaqua}\circle{34}}
 \put(-2.5,-3){\color{myaqua}\thepage}}

 \renewcommand{\headrule}{\hbox to\headwidth{\color{myaqua}\leaders\hrule height \headrulewidth\hfill}}

\sectionn{Methodology}

{ \fontfamily{times}\selectfont
 \noindent
The Nikiforov-Uvarov (NU) \cite{Nikiforov88} method was discussed in details in \cite{Farout20, Zhang10,Miranda10}. Here we will mention the main points to remind the reader with the main idea in NU method. NU is usually used to reduce the second-order differential equation, we aim to solve, into a general form of a hypergeometric type, using a suitable coordinate transformation $s=s(r)$. The form we aim to solve is:
\begin{equation}
\psi_n^{''}(s)+ \frac{\tilde{\tau}(s)}{\sigma(s)} \psi_n^{'}(s)+ \frac{\tilde{\sigma}(s)}{\sigma^2(s)} \psi_n(s)=0,
\label{eq: NU-equ}
\end{equation}
where $\sigma (s)$ and $\tilde{\sigma}(s)$ are polynomials, restricted to be at most second-degree, and $\tilde{\tau}(s)$ is required to be a first-degree polynomial. The wave function can be assumed to be of the form,
\begin{equation}
\psi_n(s)= \phi_n(s) y_n(s),
\label{eq: psi}
\end{equation}
this transforms equation (\ref{eq: NU-equ}) into a hypergeometric equation of a form
\begin{equation}
\sigma(s) y_n^{''}(s) + \tau(s) y_n^{'}(s) + \lambda y_n(s)= 0,
\label{eq:hypergeometric}
\end{equation}
where \begin{equation}
\sigma(s)= \Pi(s) \frac{\phi_n(s)}{{\phi_n}^{'}(s)},
\label{eq:pi mentioned}
\end{equation}
\begin{equation}
\tau(s)= \tilde{\tau}(s) + 2n~ \Pi(s),     
\end{equation}
and $\lambda$ is a parameter defined as,
\begin{equation}
\lambda = \lambda_n = -n \tau^{'}(s) - \frac{n(n-1)}{2} \sigma^{''}(s),
\label{eq: lambda1}
\end{equation}
$n=0, 1, 2, ......$, and $\tau(s)$ must be a polynomial with a negative first derivative to produce a solution, with a physical meaning, for equation (\ref{eq:hypergeometric}). For more details you can refer to \cite{Farout20}.

To get the eigenvalues of the system, $\lambda$ defined in equation (\ref{eq: lambda1}) can be used with 
\begin{equation}
\lambda = \lambda_n= k+ \Pi^{'}(s),
\label{eq: lambda2}
\end{equation}
}

\sectionn{Feinberg-Horodecki equation with Time-Dependent Screened Kratzer-Hellmann Potential}
\label{sec:SKHP}


{ \fontfamily{times}\selectfont
 \noindent
The NU method is used to find the approximate solutions of FH equation for the screened Kratzer-Hellmann potential, then the eigenvalues and eigenfunctions of three special cases are produced from the results.

The time-dependent screened Kratzer-Hellmann potential (SKHP) is given by \cite{Osobonye20}
\begin{equation}
V(t)= \left(\frac{V_0}{t} + \frac{V_1}{t} e^{\alpha t} + \frac{V_2}{t^2}\right) e^{- \alpha t},
\label{eq:SKH}
\end{equation}
where $V_0$ , $V_1$, and $V_2$ are adjustable real potential parameters, and $\alpha$ is dimensionless screening parameter. The time-dependent potential may represent a growing system biological or physical \cite{Molski06, Molski10}.
This potential is considered as a general case of five special cases;\\ 1) Hellmann potential when $V_2 =0$ which is defined as \cite{Hellmann35}
\begin{equation}
V(t)= \frac{V_0}{t} e^{- \alpha t} + \frac{V_1}{t} ,
\label{eq:Hell. pot}
\end{equation}
2) Screened Kratzer potential when $V_1=0$ which is defined as \cite{Ikot19}
\begin{equation}
V(t)= \left(\frac{V_0}{t} + \frac{V_2}{t^2}\right) e^{- \alpha t},
\label{eq:Screened Kratzer Pot}
\end{equation}
3) Kratzer potential when $V_1=\alpha=0$ which is defined as \cite{Kratzer20}
\begin{equation}
V(t)= \left(\frac{V_0}{t} + \frac{V_2}{t^2}\right),
\label{eq:Kratzer pot}
\end{equation}
4) Yukawa or Screened Coulomb potential when $V_1=V_2=0$, which is defined as \cite{Lam71}
\begin{equation}
V(t)= \frac{V_0}{t} e^{- \alpha t},
\label{eq:Screened coulomb pot}
\end{equation}
and 5) Coulomb potential when $V_1=V_2=\alpha=0$, which is defined as \cite{Hong78}
\begin{equation}
V(t)= \frac{V_0}{t}.
\label{eq:Coulomb pot}
\end{equation}

If the SKHP potential is substituted in FH equation, one obtains
\begin{equation}
\left[-\frac{\hbar^2}{2mc^2}\frac{d^2}{dt^2}+ \left(\frac{V_0}{t} + \frac{V_1}{t} e^{\alpha t} + \frac{V_2}{t^2}\right) e^{- \alpha t}\right]\psi_n(t) = cP_n \psi_n(t).
\end{equation}
Using the approximation \cite{Qiang07, Dong2007} defined as
\begin{equation}
\frac{1}{t} \approx \frac{\alpha}{(1-e^{-\alpha t})},
\end{equation}
then, let $s=e^{-\alpha(t)}$, where s $\gamma$(0, 1), we get
\begin{equation}
\psi_n^{''}(s) + \frac{(1-s)}{s(1-s)} \psi_n^{'}(s)+\frac{-\gamma_1^2-\gamma_3s+\gamma_2s^2}{s^2(1-s)^2}\psi_n(s)=0,
\label{eq:differential eq to solve}
\end{equation}
where 
\begin{equation}
\gamma_1^2=\frac{2mc^2}{\hbar^2 \alpha^2}\left(\alpha V_1 - cP_n\right),
\label{eq:gamma1}
\end{equation}
\begin{equation}
\gamma_2=\frac{ 2mc^2}{\hbar^2 \alpha^2}\left(\alpha V_0 +cP_n\right),
\end{equation}
\begin{equation}
\gamma_3=-\frac{2mc^2}{\hbar^2 \alpha^2}\left(\alpha V_0 - \alpha V_1+ \alpha^2 V_2 +2cP_n\right).
\end{equation}
Comparing equation (\ref{eq:differential eq to solve}) with equation (\ref{eq: NU-equ}), one obtains \\
$\tilde{\tau}(s)=1-s$, $\sigma(s)= s(1-s)$, and $\tilde{\sigma}(s)=-\gamma_1^2-\gamma_3s+\gamma_2 s^2$.\\
By substituting these values in $\Pi(s)= \frac{\sigma^{'}-\tilde{\tau}}{2}\pm \sqrt{(\frac{\sigma^{'}-\tilde{\tau}}{2})^2-\tilde{\sigma}+ k\sigma}$, one obtains
\begin{equation}
\Pi(s)= -\frac{s}{2}\pm \sqrt{\left(\frac{1}{4}-\gamma_2-k\right)s^2+ (k+\gamma_3)s+\gamma_1^2}.
\label{eq: pi solving}
\end{equation}
To find the form of $\Pi(s)$ that makes the solution has a physical meaning, the discriminant under the square root, in equation (\ref{eq: pi solving}), must be zero, so that the expression of $\Pi (s)$ becomes the square root of a polynomial of the first degree. This condition can be written as
\begin{equation}
\left(\frac{1}{4}-\gamma_2-k\right)s^2+ (k+\gamma_3)s+\gamma_1^2=0.
\end{equation}
Solving this equation, one gets
\begin{equation}
s= \frac{-(k+\gamma_3)\pm\sqrt{(k+\gamma_3)^2-4\gamma_1^2(\frac{1}{4}-\gamma_2-k)}}{2(\frac{1}{4}-\gamma_2-k)}.
\label{eq: s}
\end{equation}
Then, for our purpose we propose that
\begin{equation}
(k+\gamma_3)^2-4\gamma_1^2\left(\frac{1}{4}-\gamma_2-k\right)=0.
\end{equation}
Arranging this equation and solving it, we get an expression for k which is defined as,
\begin{equation}
k_\pm = -\gamma_3-2\gamma_1^2\pm 2\gamma_1\left(\frac{1}{\eta}-\frac{1}{2}\right),
\end{equation}
where the expression between the parentheses is given by
\begin{equation}
\frac{1}{\eta}=\frac{1}{2} +\sqrt{\frac{1}{4}+\frac{2mc^2 V_2}{\hbar^2}},
\label{eq:1/R-1/2}
\end{equation}
where $\frac{2mc^2 V_2}{\hbar^2}+ \frac{1}{4}\geq 0$.

If we substitute the form of $k_-$ into equation (\ref{eq: pi solving}) we get a possible expression for $\Pi(s)$, which is given by 
\begin{equation}
\Pi_-(s) =\gamma_1 -s( \gamma_1+\frac{1}{\eta}),
\label{eq: pi(s) solution}
\end{equation}
this solution satisfy the condition that the derivative of $\tau(s)$ is negative. Therefore, the expression of $\tau(s)$ which satisfies these conditions can be written as
\begin{equation}
\tau(s) =1-s +2\gamma_1-2s(\gamma_1+\frac{1}{\eta}).
\label{eq: tau(s) result}
\end{equation}
Now, substituting the values of $\tau^{'}_-(s)$, $\sigma^{''}(s)$, $\Pi_-(s)$ and $k_-$ into equations (\ref{eq: lambda1}) and (\ref{eq: lambda2}), we obtain
\begin{equation}
\lambda_n= \frac{-2mc^2}{\hbar^2 \alpha}\left(V_0 + V_1\right)-\frac{1}{\eta}(1+2\gamma_1),
\label{eq:lambda_n solution}
\end{equation}
and 
\begin{equation}
\lambda= \lambda_n= n^2 + 2n(\frac{1}{\eta}+ \gamma_1).
\label{eq:lambda-solution}
\end{equation}
Now, solving equations (\ref{eq:lambda_n solution}) and (\ref{eq:lambda-solution}), one obtains the eigenvalues of the quantized momentum which is defined as,
\begin{equation}
P_n= \frac{1}{c}\left[\alpha V_1 -\frac{\hbar^2 \alpha^2 }{2mc^2}\left(\frac{\frac{2mc^2}{\hbar^2 \alpha}(V_0+V_1+\alpha V_2)+n(n+\frac{2}{\eta})+\frac{1}{\eta}}{2(n+\frac{1}{\eta})}\right)^2\right].
\label{eq:eigenvalues of Pn}
\end{equation}
Due to the NU method used in getting the eigenvalues, the polynomial solutions of the hypergeometric function $y_n(s)$ depend on the weight function $\rho(s)$ which can be determined by solving $\sigma (s) \rho^{'}(s) +[\sigma(s) - \tau(s)] \rho(s) =0$ to get 
\begin{equation}
\rho(s)= s^{l_1}(1-s)^{l_2}.
\label{eq:rho-result}
\end{equation}
where $l_1=2\gamma_1$ and $l_2=(\frac{2}{\eta})-1$.

Substituting $\rho(s)$ into $y_n(s)= \frac{A_n}{\rho(s)} \frac{d^n}{ds^n} [\sigma^n(s) \rho (s)]$, we get an expression for the wave functions as
\begin{equation}
y_n(s)= A_n s^{- l_1}(1-s)^{-l_2)}\frac{d^n}{ds^n}\left[s^{n+l_1}(1-s)^{n+l_2}\right],
\label{eq: yn 2}
\end{equation}
where $A_n$ is the normalization constant. Solving equation (\ref{eq: yn 2}) gives the final form of the wave function in terms of the Jacobi polynomial $P_n^{(\alpha, \beta)}$ as follows,
\begin{equation}
y_{n}(s)= A_n n! P_n^{(l_1, l_2)}(1-2s).
\label{eq: yn last solution}
\end{equation}
Now, substituting $\Pi_-(s)$ and $\sigma(s)$ into equation (\ref{eq:pi mentioned}) then solving it we obtain
\begin{equation}
\phi_n(s)= s^{\gamma_1}(1-s)^\frac{1}{\eta}.
\label{eq: phi solutin}
\end{equation}
Substituting equations (\ref{eq: yn last solution}) and (\ref{eq: phi solutin})in equation(\ref{eq: psi}), and using $s=e^{-\alpha(t)}$ one obtains,
\begin{equation}
\psi_n(s)= B_n e^{-\alpha\gamma_1 t}(1-e^{-\alpha t})^\frac{1}{\eta} P_n^{(2\gamma_1, \frac{2}{\eta}-1)}(1-2e^{-\alpha t}),
\label{eq:psi}
\end{equation}
where $B_n$ is the normalization constant and $\frac{1}{\eta}$ is defined as in equation (\ref{eq:1/R-1/2}). 
\section{Special Cases}
\noindent 
\subsection{Time-Dependent Hellmann Potential}
\noindent 
To get the Hellman potential from the SKHP form, as mentioned above, see equation (\ref{eq:Hell. pot}). If we substitute $V_2=0$ in (\ref{eq:eigenvalues of Pn}) we get the eigenvalues of the time-dependent HF equation with Hellmann  potential. The result is as follow ,
\begin{equation}
P_n= \frac{1}{c}\left[\alpha V_1 -\frac{\hbar^2 \alpha^2 }{2mc^2}\left(\frac{\frac{2mc^2}{\hbar^2 \alpha}(V_0+V_1)+n(n+2)+1}{2(n+1)}\right)^2\right].
\end{equation}

To determine the eigenfunctions associated with the modified Kratzer potential, the same parameters were substituted in (\ref{eq:gamma1}) which results in 
\begin{equation}
\psi_n(s)= B_n e^{-\alpha \gamma_1 t} (1-e^{-\alpha t}) P_n^{(2\gamma_1, 1)}(1-2e^{-\alpha t}),
\end{equation}
where 
\begin{equation}
\gamma_1=\sqrt{\frac{2mc^2}{\alpha^2\hbar^2}} \left(\frac{\frac{2mc^2}{\hbar^2 \alpha}(V_0+V_1)+n(n+2)+1}{2(n+1)}\right),
\end{equation}
and $P_n^{(2\gamma_1, 1)}(1-2e^{-\alpha t})$ is the Jacobi polynomial.

\subsection{Time-Dependent Screened Kratzer Potential}
\noindent

As mentioned above, the screened Kratzer-Hellmann potential can be reduced to screened Kratzer by setting $V_1=0$. Therefore, by substituting $V_1=0$ in equation (\ref{eq:eigenvalues of Pn}), it gives the eigenvalues of the FH time-dependent equation with screened Kratzer potential. These eigenvalues are given by the relation,
\begin{equation}
P_n= \frac{1}{c}\left[-\frac{\hbar^2 \alpha^2 }{2mc^2}\left(\frac{\frac{2mc^2}{\hbar^2 \alpha}(V_0+\alpha V_2)+n(n+\frac{2}{\eta})+\frac{1}{\eta}}{2(n+\frac{1}{\eta})}\right)^2\right].
\label{ea:p_n of SKP}
\end{equation}
where 
\begin{equation}
\frac{1}{\eta}=\frac{1}{2} +\sqrt{\frac{1}{4}+\frac{2mc^2 V_2}{\hbar^2}}.
\label{eq:1/eta}
\end{equation}

To determine the eigenfunctions associated with the screening Kratzer potential, the same parameters were substituted in (\ref{eq:psi}) and (\ref{eq:gamma1}) which results in 
\begin{equation}
\psi_n(s)= B_n e^{-\alpha \gamma_1 t}(1-e^{-\alpha t})^\frac{1}{\eta} P_n^{(2\gamma_1, \frac{2}{\eta}-1)}(1-2e^{-\alpha t}),
\end{equation}
and 
\begin{equation}
\gamma_1=i \sqrt{\frac{2mc^3 P_n}{\alpha^2\hbar^2}} 
\end{equation}
\subsection{Time-dependent modified Kratzer potential}
\noindent The SKHP can be reduced to modified Kratzer potential by substituting $V_1=\alpha=0$ in equation (\ref{eq:SKH}) or substituting $\alpha=0$ in the screened Kratzer potential. Therefore, to get the quantized momentum of the FH equation with the modified Kratzer potential, one can substitute $\alpha=0$ in equation (\ref{ea:p_n of SKP}). The result is given by

\begin{equation}
P_n=\frac{1}{c}\left[\frac{-2mc^2}{\hbar^2}\left(\frac{2V_0}{2(n+\frac{1}{\eta})}\right)^2\right],
\end{equation}
where $\frac{1}{\eta}$ is defined as in equation (\ref{eq:1/eta}). This result is exactly the same result as in a previous work \cite{Farout20}, where $V_0$ was substituted $2 t_e D_e$.

\subsection{Time-Dependent Screened Coulomb Potential}
\noindent The SKHP can be reduced to screened Coulomb potential, as mentioned above, by substituting $V_1=V_2=0$. Therefore, by the substitution of these values in equation (\ref{eq:eigenvalues of Pn}), we obtain the eigenvalues of time-dependent FH equation with screened Coulomb potential which is given by,
\begin{equation}
P_n= \frac{1}{c}\left[-\frac{\hbar^2 \alpha^2 }{2mc^2}\left(\frac{\frac{2mc^2}{\hbar^2 \alpha}(V_0)+n(n+2)+1}{2(n+1)}\right)^2\right],
\label{eq:eigenvalues of screened coulomb}
\end{equation}
which is the same result that we got in a previous work \cite{Farout20}.
The eigenfunctions associated with the screened Coulomb potential is defined by substituting the same parameters in (\ref{eq:psi}) and (\ref{eq:gamma1}) which results in 
\begin{equation}
\psi_n(s)= B_n e^{-\alpha \gamma_1 t}(1-e^{-\alpha t}) P_n^{(2\gamma_1, 1)}(1-2e^{-\alpha t}),
\end{equation}
and 
\begin{equation}
\gamma_1=\sqrt{\frac{2mc^2}{\alpha^2\hbar^2}} \left(\frac{\frac{2mc^2}{\hbar^2 \alpha}(V_0)+n(n+2)+1}{2(n+1)}\right).
\end{equation}
If $\alpha=0$ is substituted in (\ref{eq:eigenvalues of screened coulomb}), one gets the quantized momentum for the FH equation with Coulomb potential. The result is given by
\begin{equation}
P_n= \frac{-mc^2V_0^2}{2\hbar^2(n+1)^2},
\end{equation}
which is the same result that we got in \cite{Farout20}

\sectionn{Results and discussion}

{ \fontfamily{times}\selectfont
 \noindent

\begin{table}
	\centering \caption{Spectroscopic parameters of the various diatomic molecules \cite{Okorie20, Oyewumi13}} \label{tab:Parameters}
	{\begin{tabular}{c  c   c  c }
			
		\hline\hline
		\rowcolor{lightaqua}	Molecule & $D_e$ (eV) & $t_e$ (time unit) & $\mu$ (a.m.u) \\
			\hline
			
		TiH & 2.05 &  1.781 & 0.987371   \\[2mm]
		
		ScN & 4.56 &  1.768 & 10.682771   \\[2mm]
		
		$H_2$  & 4.7446 &  0.7416 & 0.50391   \\[2mm]
		
		CuLi & 1.74 &  2.310 & 6.259494   \\[2mm]
		
		I$_2$ & 1.58179 &  2.6620 & 63.452235  \\[2mm]
		
		\end {tabular}}
		\label{table:paraeter}
\end{table}
}
\begin{table}
	\centering
	\caption{FH quantized momentum eigenvalues (eV/c) of the screened Kratzer potential (SKP) for diatomic molecules. The values of the parameters used are defined in Table \ref{table:paraeter}, and $V_0=-3$ eV, $V_1=0$ eV, $V_2= 10$ eV}
	{\begin{tabular}{p{1.5cm}p{1.5cm} c c c c }
			\hline\hline
			
			\rowcolor{lightaqua} SKP&$\alpha$&n=0&n=1&n=2&n=3\\[2mm]
			\hline
			
			\multirow{4}{12pt}{$I_2$}&	0.001& -0.382998564&-0.363610557&-0.345650694 & -0.328982080\\
			&0.01&-0.364335155 & -0.344962250&-0.327017880 &-0.310365149 \\
			&0.1& -0.203339870 & -0.185477225&-0.169082120 &-0.154017659\\
			&1.0&-1.157267601 &-1.290430915 & -1.428962341 & -1.572724983 \\[2mm]
			
			\multirow{4}{12pt}{$TiH$}&0.001& -0.386226660& -0.249403938& -0.173991739& -0.128064754\\
			&0.01&-0.370009486&-0.233302525& -0.158031152&-0.112270061  \\
			&0.1& -0.226967331&-0.101836412 &-0.040647740&-0.011476005  \\
			&1.0&-0.709504293 &-1.741977703 &-3.089059016 &-4.718822923  \\[2mm]			 
			
			\multirow{4}{12pt}{$ScN$}&	0.001&-1.072492035 &-0.961920092&-0.867548609& -0.786360699 \\[2mm]
			&0.01& -1.036586258&-0.926057079&-0.831730677&-0.750590166  \\
			&0.1& -0.711151001 & -0.604898248&-0.515079953 & -0.438679231 \\
			&1.0& -0.819049236 &-1.140439095 & -1.501431558&-1.899009739  \\[2mm]
			
			\multirow{4}{12pt}{$H_2$}&	0.001&-0.644495629&-0.303438242&-0.175331183&-0.113740718 \\
			&0.01& -0.628780486&-0.287855565 &-0.159930087 & -0.098570319 \\
			&0.1&-0.482297309 &-0.154618931 &-0.044851587 &-0.006561545   \\
			&1.0&-0.084290853 &-1.081266746 &-2.787312809 & -5.055995309 \\[2mm]
			
			\multirow{4}{12pt}{$CuLi$}&	0.001&-0.394473035&-0.332591591&-0.284143245&-0.245503715 \\
			&0.01& -0.376637860 & -0.314803045& -0.266405282& -0.227820289 \\
			&0.1&-0.220977375&-0.163805462 &-0.120466000 & -0.087334707  \\
			&1.0&  -0.933498922 & -1.342617140 &-1.805107747 & -2.317346460 \\[2mm]
			
			\hline
			\end {tabular}}
		\label{table:Numerical1}
	\end{table}
}
Now, to study the effect of the intermolecular interaction potentials on the quantized momentum states, we present a few numerical results by taking various screening parameter $\alpha$ values for five different molecules; namely, $I_2$, $TiH$, $ScN$, $H_2$ and $CuLi$.
In choosing appropriate values of potential parameters of various diatomic molecules, see Table \ref{table:paraeter},  we study the effect of time-dependent screening Kratzer potential in Feinberg-Horodecki equation. We calculate the FH quantized momentum states by changing the values of screening parameter $\alpha$. The energy of the molecules is strongly bound together and hence their corresponding quantized momentum states are shifted to the negative region as displayed in Table \ref{table:Numerical1}. These results appear to be appropriate with the screened Kratzer potential, see figure \ref{fig:SKHP}, since it is an attracting potential.  

 \fancyfoot{}
\fancyfoot[C]{\leavevmode
	\put(0,0){\color{lightaqua}\circle*{34}}
	\put(0,0){\color{myaqua}\circle{34}}
	\put(-5,-3){\color{myaqua}\thepage}}

In addition, we study the effect of time-dependent Hellmann potential in Feinberg-Horodecki equation. We compute the FH quantized momentum states by changing the values of screening parameter $\alpha$. The quantized momentum states are shifted to the positive region as given in Table \ref{table:Numerical2}. Which is coherent with a repulsive potential as shown in figure \ref{fig:SKHP}.

\begin{table}
	\centering
	\caption{FH quantized momentum eigenvalues (eV/c) of the Hellmann potential (HP) for diatomic molecules. The values of the parameters used are defined in Table \ref{table:paraeter}. Whereas, $V_0 = 3$ eV, $V_1 = 5$ eV and $V_2 = 0$ eV.}
	
		{\begin{tabular}{p{1.5cm} p{1.5cm} c c c c }
			\hline\hline
			
			\rowcolor{lightaqua} HP&$\alpha$&n=0&n=1&n=2&n=3\\[2mm]
			\hline
			
			\multirow{4}{12pt}{$I_2$}&	0.001& 0.004210728& 0.004210722&0.004210712 &0.004210698\\
			&0.01&0.042107102 & 0.042106511 & 0.042105526 & 0.042104147 \\
			&0.1& 0.421053294 & 0.420994194 & 0.420895695 & 0.420757796  \\
			&1.0&4.208759954 & 4.202849996 & 4.193000067& 4.179210167 \\[2mm]
		
			\multirow{4}{12pt}{$TiH$}&0.001& 0.003650923&0.003650544 & 0.003649911&0.003649024\\
			&0.01& 0.036497840 & 0.036459860&0.036396561 &0.036307942 \\
			&0.1& 0.363839012 & 0.360041047 & 0.353711107&  0.344849189 \\
			&1.0& 3.524451184 &3.144654734 & 2.511660652&  1.625468937 \\[2mm]

			\multirow{4}{12pt}{$ScN$}&	0.001& 0.008062068 & 0.008062033& 0.008061975&0.008061893\\
			&0.01& 0.080619630 &0.080616120 &0.080610269 & 0.080602078 \\
			&0.1& 0.806090989 & 0.805739957 & 0.805154902& 0.804335827  \\
			&1.0& 8.050378916 & 8.015275664 &7.956770245 & 7.874862657
			\\[2mm]
		
			\multirow{4}{12pt}{$H_2$}&0.001&0.003518347 & 0.003517603 & 0.003516363&0.003514626\\
			&0.01& 0.035161148 &0.035086730 &0.034962699&0.034789057   \\
			&0.1& 0.349378934 & 0.341937129 &0.329534121 &0.312169909   \\
			&1.0& 3.270535191 &2.526354682  & 1.286053835&  -0.450367352\\[2mm]
			
			\multirow{4}{12pt}{$CuLi$}&	0.001& 0.004019380&0.004019320 &0.004019220 &0.004019080\\
			&0.01&0.040192003 & 0.040186012 & 0.040176027 & 0.040162049 \\
			&0.1& 0.401740303& 0.401141213&0.400142730 & 0.398744854 \\
			&1.0& 3.999430335 & 3.939521339 &3.839673012 & 3.699885356 \\[2mm]
			\hline
			
			\end {tabular}}
			\label{table:Numerical2}
		\end{table}
	}

Finally, we study the time-dependent screened Kratzer-Hellmann oscillatory potential and investigate its behavior with changing values of screening parameter $\alpha$ as displayed in Table \ref{table:Numerical3}.
It is noteworthy to mention that when $\alpha$ takes small values, the effect of screening Kratzer is observed to be more dominant as the momentum states are seen shifted to the negative region since the molecules energy is strongly bound. However, as the values of $\alpha$ tends to rise up, the interaction between molecules decreases and the Hellmann effect is clearly apparent and thereby momentum states shift to the positive region due to repulsive dominant Hellmann potential part.

\begin{table}
	\centering
	\caption{FH quantized momentum eigenvalues (eV/c) of the screened Kratzer-Hellmann potential for diatomic molecules. The values of the parameters used are defined in Table \ref{table:paraeter}, whereas $V_0=-3$ eV, $V_1=5$ eV and $V_2=10$ eV.}
		
		{\begin{tabular}{p{1.5cm} p{1.5cm} c c c c }
			\hline\hline
			
			SKHP&$\alpha$&n=0&n=1&n=2&n=3\\[2mm]
			\hline
			
			\multirow{4}{12pt}{$I_2$}&0.001&-0.378787834  &-0.359399827 &-0.341439964 & -0.324771350\\
			&0.01&-0.322227856  & -0.302854951 & -0.284910581 & -0.268257849 \\
			&0.1& 0.217733124  & 0.235595769 & 0.251990874 &  0.267055335 \\
			&1.0& 3.053462339  &2.920299025  & 2.781767599 & 2.638004956  \\[2mm]
			
			\multirow{4}{12pt}{$TiH$}&0.001&-0.382575610 &-0.245752888 & -0.170340689 &-0.124413704 \\
			&	0.01 & -0.333498986 & -0.196792025 & -0.121520652 &  -0.075759561  \\
			& 0.1  & 0.138137669 &0.263268588 & 0.324457260 & 0.353628995  \\
			& 1.0 & 2.941545707 & 1.909072297 & 0.561990984 & -1.067772923   \\[2mm]
			\hline
			
			\multirow{4}{12pt}{$ScN$}&0.001&-1.064429955  &-0.953858012 &-0.859486529 &-0.778298619 \\
			&	0.01 & -0.955965458 & -0.845436279 & -0.751109877 & -0.669969366   \\
			& 0.1  & 0.095056999  & 0.201309752 & 0.291128047  &  0.367528769   \\
			& 1.0 &7.243030764 &  6.921640905 &  6.560648442  & 6.163070261  \\[2mm]
			\hline
			\multirow{4}{12pt}{$H_2$}&0.001&-0.640977033&-0.299919647 &-0.171812588& -0.110222123 \\
			&	0.01 &-0.593594532&-0.252669611 & -0.124744134 & -0.063384366  \\
			& 0.1  & -0.130437773 & 0.197240605 & 0.307007949 & 0.345297991   \\
			& 1.0 & 3.434304507& 2.437328614& 0.731282551 & -1.537399949   \\[2mm]
			\hline
			\multirow{4}{12pt}{$CuLi$}&0.001&-0.390453635 &-0.328572191&-0.280123845& -0.241484315 \\
			&	0.01 &-0.336443860 &-0.274609045 & -0.226211282&  -0.187626289 \\
			& 0.1  &0.180962625 & 0.238134538 & 0.281474000 & 0.314605293 \\
			& 1.0 & 3.085901078&  2.676782860 &2.214292253&  1.702053540  \\[2mm]
			\hline
			\end {tabular}}
			\label{table:Numerical3}
		
\end{table}}


\begin{figure}
	\includegraphics[width=0.99\linewidth]{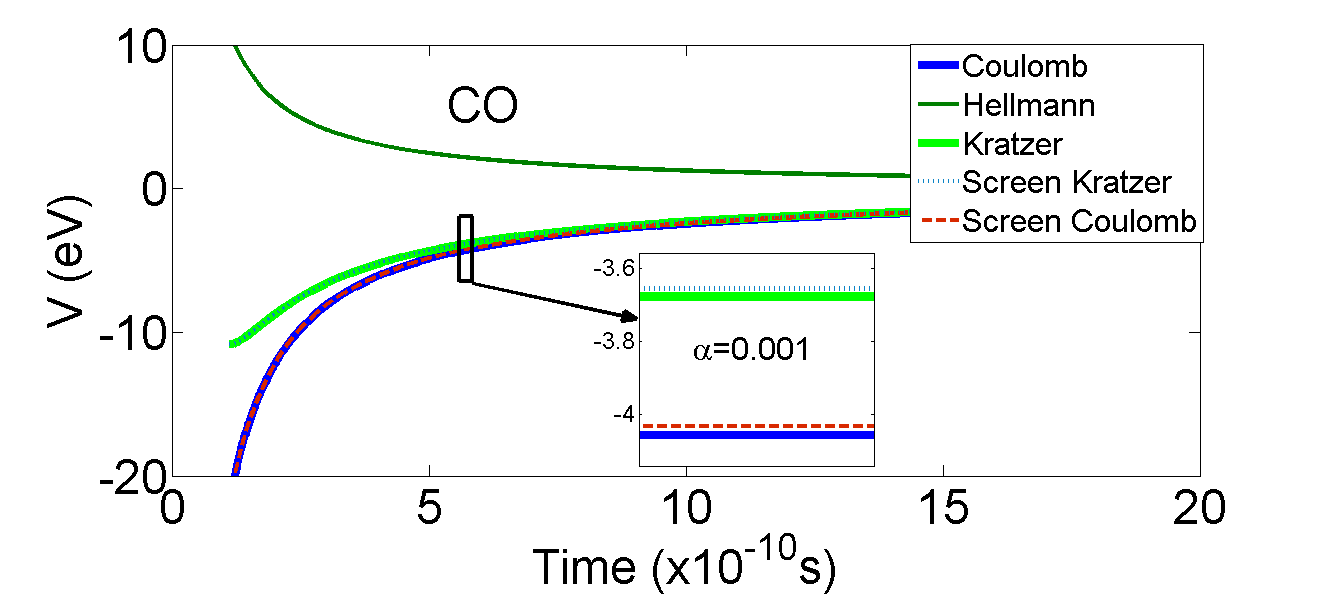}
	\caption[SKH Potential]{The screened Kratzer-Hellmann potential for diatomic molecules.}
	\label{fig:SKHP}
\end{figure}

\section{Conclusions}
\noindent We solved the Feinberg-Horodecki equation for the time-dependent screened Kratzer-Hellmann potential via Nikiforov-Uvarov method. We got the approximate quantized momentum eigenvalues solution of the FH equation.
Obviously, if one substitutes $ct=r$ and $cP_n=E_n$, the Feinberg-Horodecki equation reduces to Schrodinger equation.

This is obvious from our results, in  particular, for the time-dependent Coulomb potential where particles are strongly bound together and so the quantized momentum states. We derived this result for the oscillatory Coulomb case as well as for the oscillatory Kratzer potential in our recent work \cite{Farout20} and also in this work. In short, the energy for Coulomb and Kratzer are bound (negative) states and momentum as well.
Therefore its natural to get negative signs for momentum states as indicated in other works on FH equation. The time parameter has its importance in application of biophysics as it represents growth.
It is therefore, worth mentioning that the method is elegant and powerful. Our results can be applied in biophysics and other branches of physics.
In this paper, we have applied our result for the Hellmann, screened Kratzer, modified Kratzer, screened Coulomb and Coulomb potentials, as special cases of the used  potential, for quantized momentum eigenvalues. Obviously the momentum states are inversely proportional with square of the momentum state of the diatomic molecules. Whereas, from Coulomb case, it is obvious that the momentum eigenvalue is proportional with the mass of the diatomic molecules. Therefore, the quantized momentum is proportional with the bound state energy of states.

\vskip 3mm

 \noindent\Large\bf Acknowledgments}

 \vskip 3mm

{ \fontfamily{times}\selectfont
 \noindent
 We thank the Editor and the referees for their valuable comments.
 This research is funded by Winter School in High Energy Physics in Palestine (WISHEPP). This generous support is greatly appreciated.

 {\color{myaqua}

}}
\end{document}